\documentclass[twocolumn]{jpsj3}

\title{Molecular Dependence of the Large Seebeck Effect in $\tau$-type Organic Conductors }

\author{Hirohito Aizawa$^1$\thanks{aizawa@kanagawa-u.ac.jp}, 
Kazuhiko Kuroki$^2$, Harukazu Yoshino$^3$, George A. Mousdis$^4$, 
George C. Papavassiliou$^4$, and Keizo Murata$^3$}

\inst{
$^1$ Institute of Physics, Kanagawa University, Yokohama, Kanagawa 221-8686, Japan \\
$^2$ Department of Physics, Graduate School of Science, Osaka University, 
Toyonaka, Osaka 560-0043, Japan \\
$^3$ Graduate School of Science, Osaka City University, Osaka 558-8585, Japan \\
$^4$ Theoretical and Physical Chemistry Institute, 
National Hellenic Research Foundation, Athens 11635, Greece \\
}

\abst{ We study the Seebeck effect in the $\tau$-type organic conductors,
$\tau$-(EDO-\textit{S,S}-DMEDT-TTF)$_{2}$(AuBr$_{2}$)$_{1+y}$  
and $\tau$-(P-\textit{S,S}-DMEDT-TTF)$_{2}$(AuBr$_{2}$)$_{1+y}$, 
where EDO-\textit{S,S}-DMEDT-TTF and P-\textit{S,S}-DMEDT-TTF
are abbreviated as OOSS and NNSS, respectively, 
both experimentally and theoretically. 
Theoretically in particular, we perform first-principles 
band calculation for the two materials and construct a two-orbital model, 
on the basis of which we calculate the Seebeck coefficient. 
We show that the calculated temperature dependence of 
the Seebeck coefficient $S$ is semi-quantitatively consistent with 
the experimental observation. 
In both materials, the absolute value of the Seebeck coefficient 
is maximum at a certain temperature, 
and this temperature is lower for NNSS than for OOSS. 
From a band structure viewpoint, 
we find that this can be traced back to the narrowness of the band gap 
between the upper and the lower pudding-mold type bands. 
On the other hand, 
the Seebeck coefficient of NNSS in the low temperature regime 
steeply increases with increasing temperature, 
which is due to the narrowness of the upper band. 
These differences in thermoelectric properties demonstrate 
the effectiveness of controlling the band structure 
through molecular modification. }

\begin{document}
\maketitle

\section{Introduction}

The electronic band structure plays an important role in 
the transport properties of materials. For the Seebeck effect in particular, 
the electron-hole asymmetry in the band structure is an important factor,  
as has been shown in previous studies 
\cite{general-review,Singh-PRB-61-13397
,Wilson-Singh-PRB-75-035121,Kuroki-Arita-JPSJ-76-083707
,Arita-Kuroki-PRB-78-115121,Usui-Arita-JPCM-21-064223
,Usui-Suzuki-PRB-88-075140,Mori-Usui-JPSJ-83-023706}, 
so that tuning the band structure into ideal forms can be 
a promising way of obtaining good thermoelectric properties.
From this viewpoint, organic molecular materials are particularly 
interesting in the sense that the overlap 
between the neighboring molecular orbitals can be changed 
by either varying the lattice structure or modifying the molecules themselves, 
which in turn should affect the band structure. 
In  this kind of approach, first-principles band calculations 
are expected to be effective since such studies can guide us toward 
better band structures for thermoelectricity.

In the present paper, we focus on $\tau$-type organic conductors, 
in which a large Seebeck effect has been found 
\cite{Yoshino-Papavassiliou-JTAC, Yoshino-Aizawa-PBCM}. 
In fact, in our previous paper \cite{Aizawa-Kuroki-PBCM}, 
we studied the origin of the large Seebeck effect observed in 
$\tau$-(EDO-\textit{S,S}-DMEDT-TTF)$_{2}$(AuBr$_{2}$)$_{1+y}$, 
and showed that the pudding-mold type band \cite{Kuroki-Arita-JPSJ-76-083707}, 
introduced to understand the large Seebeck coefficient of 
Na$_x$CoO$_2$ \cite{Terasaki-Sasago-PRB-56-R12685}, plays an important role. 
There, it was also revealed that the suppression of the Seebeck 
coefficient at high temperature is due to the existence of 
a band sitting right below the band that intersects the Fermi level,
separated by a small gap. 
The temperature dependence of the Seebeck coefficient may depend on the gap, 
which in turn may be tuned by the variation of the donor molecule.

\begin{figure}[!htb]
  \centering
  \includegraphics[width=7.5cm]{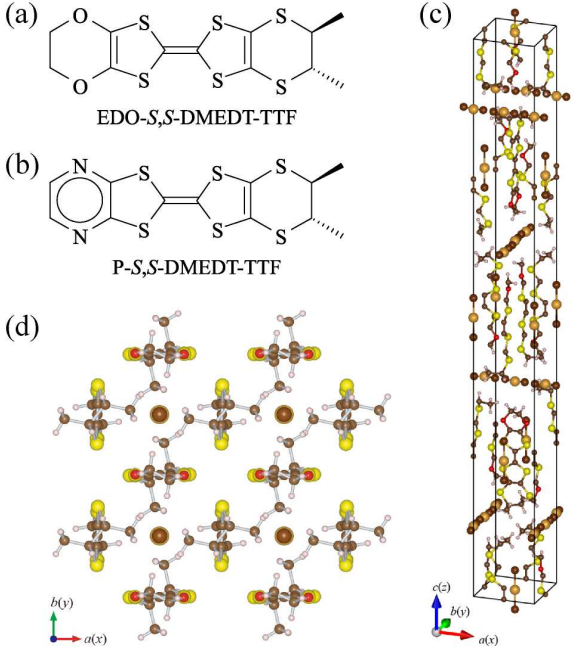}
  \caption{(Color online) 
  Molecular structures of donor molecules, 
  (a) EDO-\textit{S,S}-DMEDT-TTF abbreviated as OOSS and 
  (b) P-\textit{S,S}-DMEDT-TTF abbreviated as NNSS in this paper.  
  (c) Side view of crystal structure of $\tau$-$D_{2}$(AuBr$_{2}$)$_{1+y}$, 
  where $D$ is the OOSS or NNSS donor molecule, and 
  (d) top view of the conducting layer of $\tau$-$D_{2}$AuBr$_{2}$. 
  }
  \label{66672fig1}
 \end{figure}

Here we investigate such a possibility by studying two materials 
based on different molecules, i.e., 
$\tau$-(EDO-\textit{S,S}-DMEDT-TTF)$_{2}$(AuBr$_{2}$)$_{1+y}$  
and $\tau$-(P-\textit{S,S}-DMEDT-TTF)$_{2}$(AuBr$_{2}$)$_{1+y}$,  
where EDO-\textit{S,S}-DMEDT-TTF is an abbreviation of 
ethylenedioxy-\textit{S,S}-dimethyl(ethylenedithio)tetrathiafulvalene 
shown in Fig.  \ref{66672fig1}(a),  and 
P-\textit{S,S}-DMEDT-TTF is for 
pyrazino-\textit{S,S}-dimethyl(ethylenedithio)tetrathiafulvalene 
shown in Fig.  \ref{66672fig1}(b).
They will be abbreviated as OOSS and  NNSS, respectively, hereafter.
The side view of the crystal structure of $\tau$-$D_{2}$(AuBr$_{2}$)$_{1+y}$, 
where $D$ is the OOSS or NNSS donor molecule, is shown in Fig.  \ref{66672fig1}(c). 
The content $y$ of the anion AuBr$_{2}^{-}$ is randomly distributed 
between the conducting layers of $\tau$-$D_{2}$AuBr$_{2}$ 
shown in Fig. \ref{66672fig1}(d), 
where the horizontal donor molecules are related to 
the vertical one turning one upside down. 

We study the above two materials both experimentally and theoretically, 
and analyze the commonalities and differences between them. 
Theoretically, we particularly compare the electronic band structure 
and calculate the Seebeck coefficient using the Boltzmann's equation approach. 
The calculated results are semi-quantitatively 
in agreement with the experimental results. 
The present study shows the possibility of controlling 
thermoelectric properties by tuning the band structure through 
the modification of molecules in organic conductors.

\section{Experimental Result for the Seebeck Coefficient}

The thermoelectric powers $\Delta V$ of 
$\tau$-(NNSS)$_{2}$(AuBr$_{2}$)$_{1+y}$ and 
$\tau$-(OOSS)$_{2}$(AuBr$_{2}$)$_{1+y}$ along their 2D layer 
($\|\mib{ab}$) were measured against Cu between 4.2 K and 300 K.  
The maximum temperature difference between two contacts of each sample 
and two Cu leads was typically 
about 0.5 K below 20 K and about 1.5 K below 300 K.  
The temperature of one of the contacts and the 
temperature difference $\Delta T$ were determined by combining two 
Chromel-Au+0.07\%Fe thermocouples and a resistance thermometer while 
the temperature difference was changed in a stepwise manner.  
Seebeck coefficient $S$ was determined as the slope of a line fit to a plot 
of $\Delta V$ vs $\Delta T$.  Finally, the absolute value of $S$ of 
the sample was obtained by adding that of Cu at each temperature.  
The experimental methods are described in detail in literatures 
\cite{Saito-JPSJ-58-4093, Saito-JPSJ-62-1001}.

\begin{figure}[!htb]
  \centering
  \includegraphics[width=7.5cm]{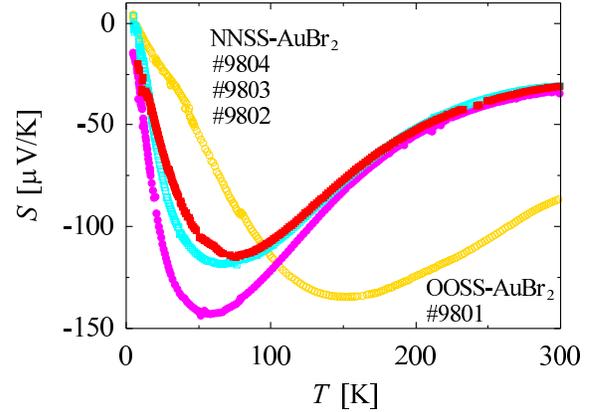}
  \caption{(Color online) 
  Experimental result of 
  the temperature dependence of the Seebeck coefficient 
  for $\tau$-(NNSS)$_{2}$(AuBr$_{2}$)$_{1+y}$ 
  and $\tau$-(OOSS)$_{2}$(AuBr$_{2}$)$_{1+y}$. }
  \label{66672fig2}
 \end{figure}

In Fig.  \ref{66672fig2}, we show the experimental results
for the temperature dependence of $S$ 
of $\tau$-(NNSS)$_{2}$(AuBr$_{2}$)$_{1+y}$ and 
$\tau$-(OOSS)$_{2}$(AuBr$_{2}$)$_{1+y}$. 
We measured $S$ of $\tau$-(NNSS)$_{2}$(AuBr$_{2}$)$_{1+y}$ for 
three samples (\#9802, \#9803, and \#9804) 
since its $S$ showed a larger sample dependence below 200 K. 
This is probably attributed to the change 
in the composition $y$ from sample to sample.
In both materials, the absolute value of the Seebeck coefficient 
first increases with increasing temperature,
reaches a maximum at a certain temperature, 
and then decreases at higher temperature. 
The temperature at which $|S|$ is maximized is 
significantly lower for the NNSS salt than for the OOSS one. 
The maximum value of $|S|$ itself is about the same between the two materials.

\section{Theoretical Method}
\label{Method}

We perform first-principles band structure calculation 
using all-electron full potential linearized augmented plane-wave (LAPW) 
+ local orbitals (lo) method to solve the Kohn-Sham equation  
using density functional theory (DFT) 
within the framework of WIEN2k \cite{WIEN2k}. 
The exchange correlation potential is calculated 
using the generalized gradient approximation (GGA). 
In the actual material, 
the content $y$ of the anion is  randomly distributed 
between the conducting layers of the $\tau$-$D_{2}$AuBr$_{2}$, 
as shown in Fig. \ref{66672fig1}(c). 
Here, we ignore this effect for simplicity in the band structure calculation, 
but adopt the experimentally determined lattice parameters of 
$\tau$-$D_{2}$(AuBr$_{2}$)$_{1+y}$. 

The single-particle wave functions in the interstitial region are 
expanded by plane waves with a cut-off of $R_{\rm MT} K_{\rm max}=3.0$,  
where $R_{\rm MT}$ denotes the smallest muffin tin radius 
and $K_{\rm max}$ is the maximum value of the wave vector 
in the plane wave expansion. 
In $\tau$-(OOSS)$_{2}$AuBr$_{2}$, 
the muffin-tin radii are assumed to 
be 2.38, 2.11, 1.61, 1.27, 1.18, and 0.64 atomic units (au) 
for Au, Br, S, O, C, and H, respectively. 
In $\tau$-(NNSS)$_{2}$AuBr$_{2}$, 
the muffin-tin radii are assumed to 
be 2.38, 2.11, 1.62, 1.24, 1.12, and 0.60 au 
for Au, Br, S, N, C, and H, respectively. 
$K_{\rm max}$ is taken as 4.7, and 
the plane wave cutoff energy is 298.8 eV for $\tau$-(OOSS)$_{2}$AuBr$_{2}$. 
In $\tau$-(NNSS)$_{2}$AuBr$_{2}$, 
$K_{\rm max}$ is taken as 5.0, and the plane wave cutoff energy is 340.0 eV. 
The wave functions in the muffin tin spheres are 
expanded up to $l_{\rm max} = 10$, 
while the charge density was Fourier expanded up to 
$G_{\rm max}=20$. 
Calculations are performed using 8$\times$8$\times$8 $k$-points 
in the irreducible Brillouin zone. 

\begin{figure}[!htb]
  \centering
  \includegraphics[width=7.5cm]{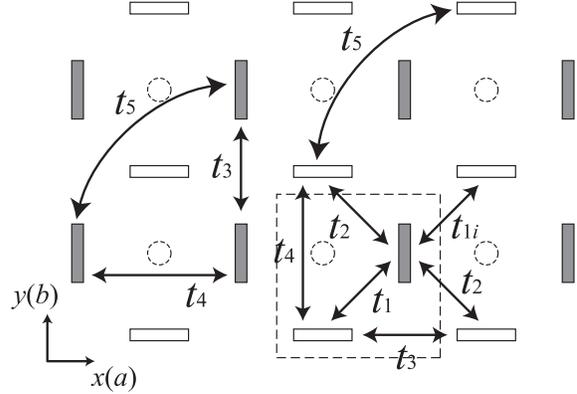}
  \caption{ Tight binding model for $\tau$-type organic conductors, 
  $\tau$-(OOSS)$_{2}$AuBr$_{2}$ and $\tau$-(NNSS)$_{2}$AuBr$_{2}$. 
  The open and filled rectangles represent the donor molecules 
  turning one upside down, respectively. 
  The dashed circles represent the anions in the conducting layer. 
  }
  \label{66672fig3}
 \end{figure}

Since there are two donor molecules having different orientations, 
as shown in Fig. \ref{66672fig1}(d), 
we obtain the tight binding model with two sites in a unit cell 
by fitting the first-principles band structure.
The tight binding model shown in Fig.  \ref{66672fig3} is 
\begin{eqnarray}
 H=\sum_{\left< i \alpha; j \beta \right>, \sigma}
  \left\{ t_{i \alpha; j \beta} 
   c_{i \alpha \sigma}^{\dagger} c_{j \beta \sigma} + {\rm H. c.} 
  \right\}, 
\label{Hij}
\end{eqnarray} 
where 
$i$ and $j$ are indices of the unit cells, 
$\alpha$ and $\beta$ are indices of the molecules (=sites) in a unit cell, 
$c_{i \alpha \sigma}^{\dagger}$ denotes the creation operator of electrons 
with spin $\sigma$ at the $\alpha$ molecule in the $i$-th unit cell, 
and $t_{i \alpha; j \beta}$ is the transfer energy 
between sites $(i, \alpha)$ and $(j, \beta)$.
Here, $t_1=t_{1i}$ holds by symmetry for the actual $\tau$-$D_{2}$AuBr$_{2}$, 
but we assign different notations 
to consider the effect of tuning  the band structure 
through chemical modification later. 
Eq. (\ref{Hij}) can be Fourier transformed as 
\begin{eqnarray}
 H=\sum_{\textbf{\textit k}, \sigma, \alpha, \beta}
  \varepsilon_{\alpha \beta}\left( \textbf{\textit{k}} \right)
   c_{\textbf{\textit k} \alpha \sigma}^{\dagger} 
   c_{\textbf{\textit k} \beta \sigma}, 
\end{eqnarray}
where $\varepsilon_{\alpha \beta}\left( \textbf{\textit{k}} \right)$ are 
\begin{eqnarray}
 \varepsilon_{11}\left( \textbf{\textit{k}} \right)
  &=& 2t_{3} \cos\left( k_{x} \right)+2t_{4} \cos\left( k_{y} \right)
\nonumber \\
 & &  +4t_{5} \cos\left( k_{x} \right) \cos\left( k_{y} \right), 
\\
  \varepsilon_{22}\left( \textbf{\textit{k}} \right)
  &=& 2t_{3} \cos\left( k_{y} \right)+2t_{4} \cos\left( k_{x} \right)
\nonumber \\
 & &  +4t_{5} \cos\left( k_{x} \right) \cos\left( k_{y} \right), 
\\
  \varepsilon_{12}\left( \textbf{\textit{k}} \right)
  &=& t_{1}+t_{2}e^{-ik_{x}}+t_{2}e^{-ik_{y}}
     +t_{1i}e^{-i(k_{x}+k_{y})}, 
\,\,\,\,\,\,
\end{eqnarray}
and $\varepsilon_{21}\left( \textbf{\textit{k}} \right) 
= \varepsilon_{12}^{*}\left( \textbf{\textit{k}} \right)$ is satisfied.

Using the Boltzmann's equation approach, 
the Seebeck coefficient is given as 
\begin{eqnarray}
 \textbf{\textit S} 
 = \frac{1}{eT} \textbf{\textit K}_{0}^{-1} \textbf{\textit K}_{1}, 
 \label{Seebeck}
\end{eqnarray}
where $e (< 0)$ is the electron charge and $T$ is the temperature. 
$\textbf{\textit K}_{0}$ and $\textbf{\textit K}_{1}$ are 
tensors given by 
\begin{eqnarray}
 \textbf{\textit K}_{n}=\sum_{\textbf{\textit k}}
  \tau\left( \textbf{\textit k} \right)
  \textbf{\textit v}\left( \textbf{\textit k} \right)
  \textbf{\textit v}\left( \textbf{\textit k} \right)
  \left\{ 
   -\frac{\partial f\left( \varepsilon_{\textbf{\textit k}} \right)}
   {\partial \varepsilon_{\textbf{\textit k}} }
  \right\}
  \left( \varepsilon_{\textbf{\textit k}} - \mu \right)^{n}, 
 \label{Kn}
\end{eqnarray}
where $\tau\left( \textbf{\textit k} \right)$ is 
the quasiparticle lifetime, 
$\varepsilon_{\textbf{\textit k}}$ is the band dispersion, 
$\textbf{\textit v}\left( \textbf{\textit k} \right) = 
\nabla_{\textbf{\textit k}} 
\varepsilon_{\textbf{\textit k}}$ 
is the group velocity, 
$f\left( \varepsilon \right)$ is the Fermi distribution function, 
and $\mu$ is the chemical potential. 
Although the quasiparticle lifetime 
$\tau$ is generally a function of $\textbf{\textit k}$, 
in the present study, we take it as a constant,  
so that it cancels out between the denominator and the 
numerator in Eq. (\ref{Seebeck}).
In the actual calculation, 
we use the band dispersion $\varepsilon_{\textbf{\textit k}}$ 
of the obtained tight binding model.

\section{Calculation Results}

\subsection{Band structure}

\begin{figure}[!htb]
  \centering
 \includegraphics[width=7.5cm]{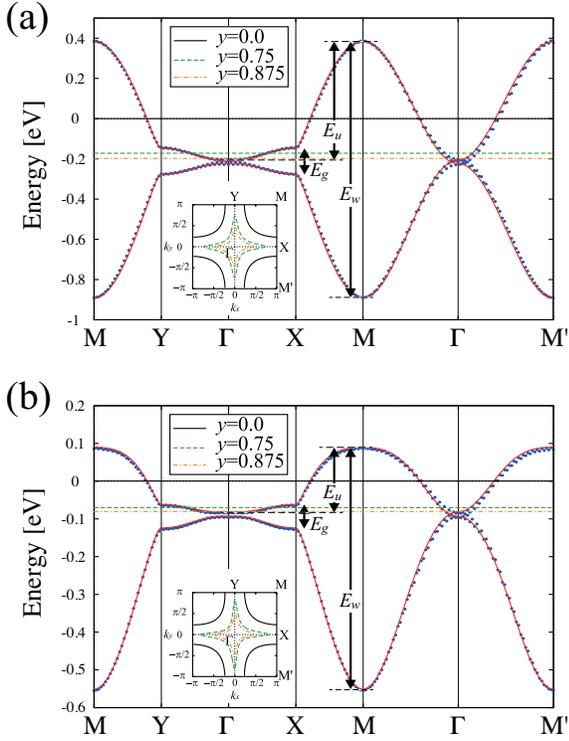}
  \caption{(Color online) 
 Calculated first-principles band structure for 
 (a) $\tau$-(OOSS)$_{2}$AuBr$_{2}$ and 
 (b) $\tau$-(NNSS)$_{2}$AuBr$_{2}$ 
 are represented by blue dotted curves. 
 The red solid lines are the tight binding model 
 fit to the first-principles band. 
 The insets show the Fermi surfaces of 
 the obtained tight binding model for each salt 
 with $y=0.0, 0.75, 0.875$, respectively. }
 \label{66672fig4}
\end{figure}

The calculated band structures of $\tau$-(OOSS)$_{2}$AuBr$_{2}$ and 
$\tau$-(NNSS)$_{2}$AuBr$_{2}$ are shown in Fig. \ref{66672fig4}. 
The two bands near the Fermi level $(E=0)$ are isolated from 
the other bands, so that extracting them using 
a tight binding model can be considered as appropriate. 

The upper and lower bands near the Fermi level are separated 
by a small band gap. 
Although the band-width of the upper band is comparable to that of 
the lower band in $\tau$-(OOSS)$_{2}$AuBr$_{2}$, 
the upper band is narrower than the lower band in 
$\tau$-(NNSS)$_{2}$AuBr$_{2}$. 
The total band-width from the top of the upper band 
to the bottom of the lower band ($E_{w}$) is estimated 
as 1.28 eV for $\tau$-(OOSS)$_{2}$AuBr$_{2}$, 
which is larger than that of $\tau$-(NNSS)$_{2}$AuBr$_{2}$, 0.65 eV. 
The upper band width ($E_{u}$) is estimated 
as 0.59 eV for $\tau$-(OOSS)$_{2}$AuBr$_{2}$, 
which is approximately three times larger than 
that of $\tau$-(NNSS)$_{2}$AuBr$_{2}$, 0.17 eV. 
As for the maximum band gap between the bottom of the upper band 
and the top of the lower band ($E_{g}$), 
it is estimated as 0.13 eV for $\tau$-(OOSS)$_{2}$AuBr$_{2}$, 
which is larger than that for 
$\tau$-(NNSS)$_{2}$AuBr$_{2}$, estimated as 0.06 eV. 

Note that the first-principles calculation results of 
the bands near the Fermi level are basically similar to the result of 
the extended H$\ddot{\rm u}$ckel calculation method, 
but the band gap between the bottom of the upper band 
and the top of the lower band is somewhat dispersive compared with 
that of the extended H$\ddot{\rm u}$ckel result 
\cite{Papavassiliou-Lagouvardos-MCLC-285-83}. 

The red solid curves in Fig. \ref{66672fig4} indicate 
the band structure of the tight binding model 
(Fig. \ref{66672fig3}) fit to the first-principles band. 
We list the transfer and important energies in Table \ref{tab1}. 
We can see that the original first-principles band structures 
are well reproduced by adopting these transfer energies. 

\begin{table}[htbp]
 \begin{center}
  \caption{ 
  List of transfer energies in the two-band 
  tight binding model shown in Fig. \ref{66672fig3} 
  and important energies 
  for $\tau$-(OOSS)$_{2}$AuBr$_{2}$ and $\tau$-(NNSS)$_{2}$AuBr$_{2}$. }
  \label{tab1}
  \begin{tabular}{lcc}
   \hline 
   \phantom{ 2} [eV]    \phantom{1} & 
   \phantom{22} OOSS \phantom{2} & 
   \phantom{22} NNSS \phantom{2}  \\ \hline\hline
   $t_{1}= t_{1i}$ & \phantom{-}0.1616           & \phantom{-}0.0817           \\ \hline
   $t_{2}$         &           -0.1573           &           -0.0787           \\ \hline
   $t_{3}$         &           -0.0114           & \phantom{-}0.0258           \\ \hline
   $t_{4}$         & \phantom{-}0.0213           & \phantom{-}0.0101           \\ \hline
   $t_{5}$         &           -0.0028           &           -0.0083           \\ \hline
\hline
   $E_{w}$         & \phantom{-}1.28\phantom{22} & \phantom{-}0.65\phantom{22} \\ \hline
   $E_{u}$         & \phantom{-}0.59\phantom{22} & \phantom{-}0.17\phantom{22} \\ \hline
   $E_{g}$         & \phantom{-}0.13\phantom{22} & \phantom{-}0.06\phantom{22} \\ \hline
  \end{tabular} 
 \end{center}
\end{table}

\subsection{Seebeck coefficient}

\begin{figure}[!htb]
  \centering
  \includegraphics[width=7.5cm]{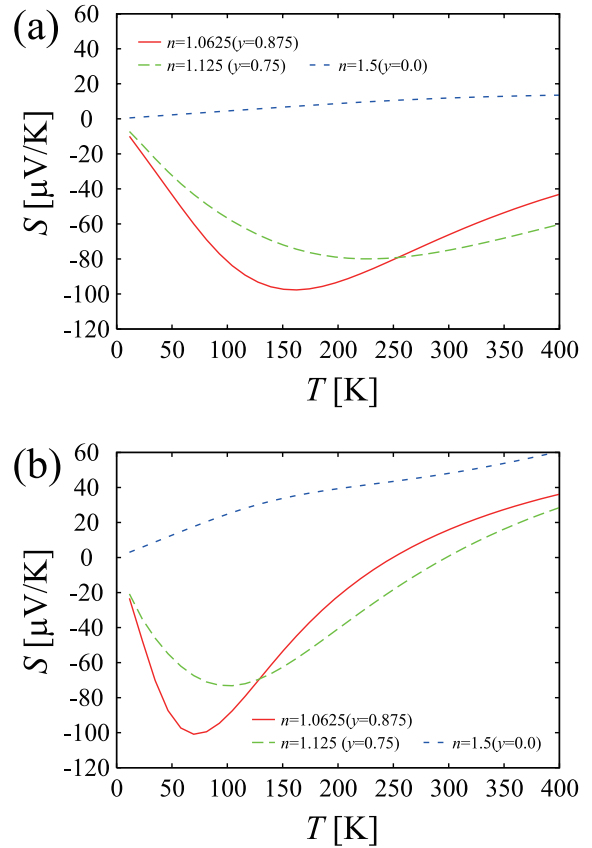}
  \caption{(Color online) 
  Calculated Seebeck coefficient $S$ in the model 
  for (a) $\tau$-(OOSS)$_{2}$(AuBr$_{2}$)$_{1+y}$ 
  and (b) $\tau$-(NNSS)$_{2}$(AuBr$_{2}$)$_{1+y}$. 
  Here, the red solid curve represents the result in $y=0.875$ ($n=1.0625$), 
  the green dashed curve is in $y=0.75$ ($n=1.125$), and 
  the blue dotted curve is in $y=0.0$ ($n=1.5$), respectively. }
  \label{66672fig5}
 \end{figure}

Figures \ref{66672fig5}(a) and \ref{66672fig5}(b) 
show the calculated Seebeck coefficient $S_{(xx)}$ 
for $\tau$-(OOSS)$_{2}$(AuBr$_{2}$)$_{1+y}$ and 
$\tau$-(NNSS)$_{2}$(AuBr$_{2}$)$_{1+y}$, respectively,  
for several band fillings.
The relation between the electron band filling $n$ and $y$ 
is represented as $n=(3-y)/2$. 
Although the precise anion content $y$ of the actual materials is not known, 
it is actually finite so that $n$ becomes smaller than 1.5 
\cite{Yoshino-JPSJ-74-417}. 
In both materials, 
the maximum absolute value of the Seebeck coefficient $\left| S \right|$ 
becomes larger as $y$ is increased ($n$ is decreased), 
and the temperature at which $\left| S \right|$ takes its 
maximum becomes lower. 
As for the comparison between the two materials, the temperature at 
which $\left| S \right|$ is maximized is lower for NNSS than for OOSS, 
which is semi-quantitatively consistent with the experimental observation.

\begin{figure}[!htb]
  \centering
  \includegraphics[width=7.5cm]{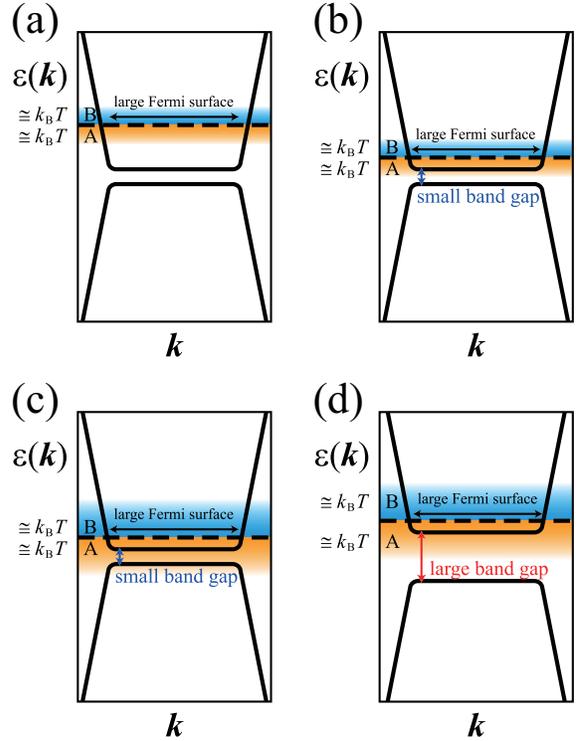}
  \caption{(Color online) 
  Schematic figure for the Fermi level and band structure 
  of $\tau$-type organic conductors, 
  where Region A  (B) is the energy range within $\simeq k_{\rm B}T$ 
  below (above) the Fermi energy. 
  (a) Small $y$, small gap, low $T$, 
  (b) large $y$, small gap, low $T$, 
  (c) large $y$, small gap, high $T$, and 
  (d) large $y$, large gap, high $T$.
} 
  \label{66672fig6}
 \end{figure}

These calculation results can be understood as follows 
on the basis of the band structure. 
As mentioned in our previous study\cite{Aizawa-Kuroki-PBCM}, 
the band structure of $\tau$-type organic conductors 
can be considered as a combination of two oppositely placed 
pudding-mold type bands composed of flat and dispersive portions,  
as schematically shown in Fig. \ref{66672fig6}. 
In region A (B), holes (electrons) are excited at finite temperature. 
When the anion content $y$ is small and the band filling is large, 
the Fermi level is located far above the band bottom of the upper band, 
so that the typical group velocities of holes and electrons are 
almost similar as shown in Fig. \ref{66672fig6}(a). 
Then the Seebeck coefficient becomes small. 
When the anion content $y$ is increased, as shown in Fig. \ref{66672fig6}(b), 
the typical group velocities of holes and electrons are 
largely different in the pudding-mold type band, 
which gives a large Seebeck coefficient 
in the low-temperature regime. 
On the other hand,  as the temperature increases, 
the holes in the lower band begin to 
contribute to the Seebeck coefficient, as shown in Fig. \ref{66672fig6}(c). 
This suppresses the Seebeck coefficient, 
so that there is a temperature at which $\left| S \right|$ is maximized. 
This temperature is lower for 
(i) a larger $y$ (smaller band filling),  
as can be understood from Figs. \ref{66672fig6}(a)-(c), and also for 
(ii) materials with a smaller band gap, 
as can be understood from the comparison 
between Figs. \ref{66672fig6}(c) and (d). 
Since the maximum band gap is smaller for NNSS, 
the temperature at which $\left| S \right|$ is maximized is lower 
when compared at the same anion content $y$.
Although the anion content of the actual materials is not well known,
the comparison between the experiment (Fig. \ref{66672fig2}) 
and the theoretical calculation (Fig. \ref{66672fig5}) 
tells us that (i) the anion content of the actual OOSS and NNSS is 
close since the experimentally observed 
maximum $\left| S \right|$ values are about the same between the two materials, 
and (ii) there is a slight variance of the anion content among different 
samples of NNSS because the maximum $\left| S \right|$ is reduced as the 
temperature at which $\left| S \right|$ is maximized becomes higher.

The above argument shows that the absolute value of 
the Seebeck coefficient takes its maximum at a lower temperature 
in NNSS because of the narrowness of the band gap $E_{g}$ in Table \ref{tab1}, 
which is a flaw of this material from the viewpoint of thermoelectric properties. 
On the other hand, the steep increase in the Seebeck coefficient 
for NNSS with increasing $T$ at low temperatures, 
as compared with that for OOSS, 
is due to the narrowness of the upper band $E_{u}$ in this material.

\begin{figure}[!htb]
  \centering
  \includegraphics[width=7.5cm]{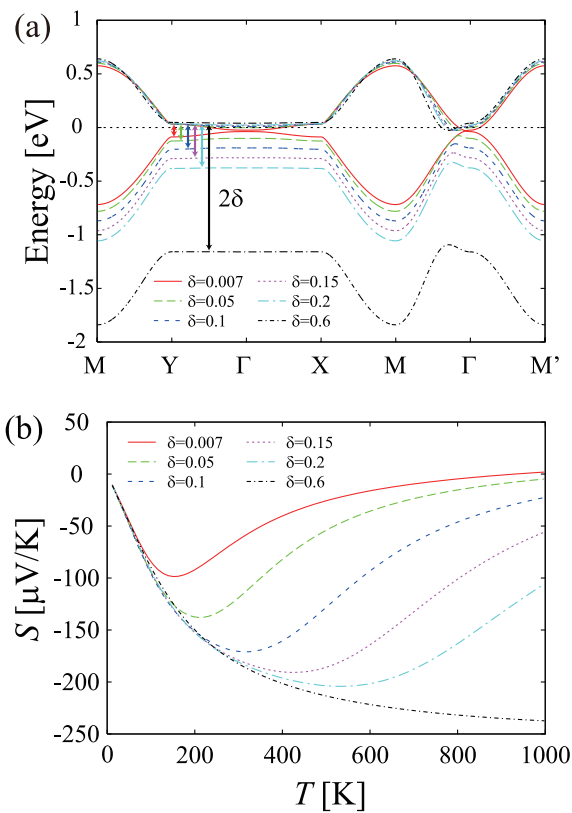}
  \caption{(Color online) 
  (a) Band structure for various values of the parameter $\delta$ 
  in the modified model. 
  The Fermi energy here is for the electron  band filling $n=1.0625$. 
  (b) Calculated Seebeck coefficient $S$ as a function of temperature 
  in the modified model.  }
  \label{66672fig7}
 \end{figure}

Therefore, if the band gap can be increased 
without affecting the band width of the upper band, 
the maximum Seebeck coefficient is expected to be larger. 
Within the present model, this can be accomplished 
by modifying the transfer energies as 
$t_{1}=0.16+\delta$, $t_{1i}=0.16$, $t_{2}=-0.16$, 
$t_{3}=-0.01$, $t_{4}=0.02$, and $t_{5}=-0.003$ eV. 
Figure \ref{66672fig7}(a) shows the variance of the band structure 
with increasing $\delta$. 
Increasing $\delta$ enlarges the band gap ($2\delta$), 
while keeping the shape and width of each band essentially unchanged.
The Seebeck coefficient calculated for 
this model is shown in Fig. \ref{66672fig7}(b). 
As expected, the maximum value of $\left| S \right|$ 
increases as  $\delta$ is increased. 
At present, we do not know a specific way of introducing 
this kind of modulation in the transfer energies. 
A finite $\delta$ is naturally introduced 
by removing the two-fold symmetry of the donor molecule around its long axis. 
For example, 
changing the configuration of the two methyl groups from ($S$, $S$) to ($S$, $R$) 
meets the purpose as long as the $\tau$-type structure is intact. 
This issue remains as an interesting subject for a future study.

\section{Conclusions}
\label{Conclusion}

We have studied the Seebeck effect in $\tau$-type organic conductors 
both experimentally and theoretically. 
We have shown that the calculated temperature dependence of 
the Seebeck coefficient is semi-quantitatively consistent 
with the experimental observation. 
In both materials, the absolute value of the Seebeck coefficient 
takes its maximum at a certain temperature, 
and this temperature is lower for NNSS than for OOSS. 
On the basis of band structure, 
this can be traced back to the narrowness of the band gap 
between the upper and lower pudding-mold type bands. 
On the other hand, the Seebeck coefficient of NNSS 
in the low-temperature regime steeply increases with increasing temperature, 
which is due to the narrowness of the upper band. 

These differences in the thermoelectric properties demonstrate 
the effectiveness of controlling the band structure 
through molecular modification. 
If the transfer energies between the molecules can be modified 
in a manner that affects only the band gap, 
even better thermoelectric properties are expected. 

In the present study, 
electron correlation effects beyond GGA are not taken into account. 
On the other hand, 
it has been known that electron correlation effects can 
in general affect the transport properties
\cite{Kontani,Kuroki-Arita-JPSJ-76-083707}.
Specifically for the present material, 
a weak ferromagnetism has in fact been experimentally suggested 
in the $\tau$-type organic conductors of the OOSS molecule 
\cite{Yoshino-JPSJ-68-117, Konoike-SM-120-801, Nakanishi-JLTP-142-247}. 
A theoretical work on the single-band Hubbard model of this material 
has shown that ferromagnetism can appear originating from 
a flat-bottomed band dispersion and a dilute band filling
\cite{Arita-PRB-61-3207}. 
Also, recent theoretical work on the spin and charge susceptibilities 
of the two-band model has shown that 
both susceptibilities are maximized at the nesting vector such as 
$\textbf{Q}=(0, 0)$, which may be related to some uniform ordering state 
\cite{Aizawa-PSS-9-1196}. 
On the other hand, 
the fact that the present calculation result of the Seebeck coefficient 
is semi-quantitatively consistent with the experiment suggests that 
the above-mentioned electron correlation effects do not play 
a significant role as far as the Seebeck effect of 
the $\tau$-type organic conductors is concerned. 
Still, the calculated result is somewhat smaller than the experimental result, 
which may be due to the correlation effects that are not taken into account, 
or to the constant $\tau$ approximation adopted here.

\section*{Acknowledgments}
This work was supported by Grants-in-Aid for Scientific Research 
from the Ministry of Education, Culture, Sports, Science and Technology of
Japan, and from the Japan Society for the Promotion of Science 
(Nos.20110007 and 26610101).


\begin{thebibliography}{99} 



\bibitem{general-review}
For a general review of the theoretical aspects 
as well as experimental observations of thermopower, 
see G. D. Mahan, Solid State Phys. \textbf{51}, 81 (1997). 

\bibitem{Singh-PRB-61-13397}
D. J. Singh, 
Phys. Rev. B  \textbf{61}, 13397 (2000). 

\bibitem{Wilson-Singh-PRB-75-035121}
G. B. Wilson-Short, D. J. Singh, M. Fornari, and M. Suewattana, 
Phys. Rev. B \textbf{75}, 035121 (2007). 

\bibitem{Kuroki-Arita-JPSJ-76-083707}
K. Kuroki and R. Arita, 
J. Phys. Soc. Jpn.  \textbf{76}, 083707 (2007). 

\bibitem{Arita-Kuroki-PRB-78-115121}
R. Arita, K. Kuroki, K. Held, A. V. Lukoyanov, S. Skornyakov, and V. I. Anisimov, 
Phys. Rev. B \textbf{78}, 115121 (2008). 

\bibitem{Usui-Arita-JPCM-21-064223}
H. Usui, R. Arita, and K. Kuroki, 
J. Phys.: Condens. Matter \textbf{21}, 064223 (2009).

\bibitem{Usui-Suzuki-PRB-88-075140}
H. Usui, K. Suzuki, K. Kuroki, S. Nakano, K. Kudo, and M. Nohara, 
Phys. Rev. B \textbf{88}, 075140 (2013). 

\bibitem{Mori-Usui-JPSJ-83-023706}
K. Mori, H. Usui, H. Sakakibara, and K. Kuroki, 
J. Phys. Soc. Jpn. \textbf{83}, 023706 (2014). 


\bibitem{Yoshino-Papavassiliou-JTAC} 
H. Yoshino, G. C. Papavassiliou, and K. Murata, 
J. Therm. Anal. Cal. \textbf{92}, 457 (2008). 

\bibitem{Yoshino-Aizawa-PBCM} 
H. Yoshino, H. Aizawa, K. Kuroki, G. C. Anyfantis, G. C. Papavassiliou, and K. Murata, 
Physica B: Condens. Matt. \textbf{405}, S79 (2010). 

\bibitem{Aizawa-Kuroki-PBCM} 
H. Aizawa, K. Kuroki, H. Yoshino, and K. Murata, 
Physica B: Condens. Matt. \textbf{405}, S27 (2010). 

\bibitem{Terasaki-Sasago-PRB-56-R12685} 
I. Terasaki, Y. Sasago, and K. Uchinokura, 
Phys. Rev.  B \textbf{56}, R12685 (1997). 

\bibitem{Saito-JPSJ-58-4093}
K. Saito, H. Kamio, Y. Honda, K. Kikuchi, K. Kobayashi, and I. Ikemoto, 
J. Phys. Soc. Jpn.  \textbf{58}, 4093 (1989). 


\bibitem{Saito-JPSJ-62-1001}
K. Saito, H. Yoshino, K. Kikuchi, K. Kobayashi, and I. Ikemoto, 
J. Phys. Soc. Jpn.  \textbf{62}, 1001 (1993). 


\bibitem{WIEN2k}
P. Blaha, K. Schwarz, G. K. H. Madsen, D. Kvasnicka, and J. Luitz, 
In WIEN2k, An Augmented Plane Wave + Local Orbitals Program 
for Calculating Crystal Properties, 
(Karlheinz Schwarz/ Techn. Universit$\ddot{\rm a}$t Wien, 
Wien, Austria, 2001). 

\bibitem{Papavassiliou-Lagouvardos-MCLC-285-83}
G. C. Papavassiliou, D. J. Lagouvardos, J. S. Zambounis, A. Terzis, 
C. P. Raptopoulou, K. Murata, N. Shirakawa, L. Ducasse, and P. Delhaes, 
Mol. Cryst. Liq. Cryst. \textbf{285}, 83 (1996). 

\bibitem{Yoshino-JPSJ-74-417}
H. Yoshino, K. Murata, T. Nakanishi, L. Li, 
E. S. Choi, D. Graf, J. S. Brooks, Y. Nogami, and G. C. Papavassiliou, 
J. Phys. Soc. Jpn. \textbf{74}, 417 (2005). 


\bibitem{Kontani}
H. Kontani, K. Kanki, and K. Ueda, Phys. Rev. B {\bf 59}, 14723 (1999).

\bibitem{Yoshino-JPSJ-68-117}
H. Yoshino, K. Iimura, T. Sasaki, A. Oda, G. C. Papavassiliou, and K. Murata, 
J. Phys. Soc. Jpn. \textbf{68}, 177 (1999).

\bibitem{Konoike-SM-120-801}
T. Konoike, A. Oda, K. Iwashita, T. Yamamoto, H. Tajima, H. Yoshino, 
K. Ueda, T. Sugimoto, K. Hiraki, T. Takahashi, T. Sasaki, Y. Nishio, 
K. Kajita, G. C. Papavassiliou, G. A. Mousdis, and K. Murata, 
Synth. Met. \textbf{120}, 801 (2001).

\bibitem{Nakanishi-JLTP-142-247}
T. Nakanishi, S. Yasuzuka, M. Teramura, L. Li, T. Fujimoto, T. Sasaki, 
T. Konoike, T. Terashima, S. Uji, Y. Nogami, G. C. Anyfantis, 
G. C. Papavassiliou, and K. Murata, 
J. Low Temp. Phys. \textbf{142}, 247 (2006).

\bibitem{Arita-PRB-61-3207}
R. Arita, K. Kuroki, and H. Aoki, 
Phys. Rev. B \textbf{61}, 3207 (2000). 

\bibitem{Aizawa-PSS-9-1196}
H. Aizawa and K. Kuroki, 
Phys. Status Solidi C \textbf{9}, 1196 (2012). 


\end{thebibliography}
\end{document}